\documentclass[aps,pre]{revtex4}
\usepackage{amssymb}
\usepackage{amsfonts}
\usepackage{amsmath}
\usepackage{graphicx}
\usepackage{epstopdf}
\usepackage{pstricks}
\usepackage{color}
\usepackage{hyperref}
\usepackage{pifont}
\usepackage{bm}
\usepackage{wasysym}

\def \Ca  {\mbox{Ca}}
\def \gdot {\dot{\gamma}}
\def \mue {\mu_{\rm e}}
\def \muf {\mu_0}
\def \mum {\mu_m}
\def \phie {\Phi_{\rm e}}

\newcommand{\Eq}[1]{Eq.~(\ref{#1})}
\newcommand{\Fig}[1]{Fig.~(\ref{#1})}

\newcommand{\ie}{i.e.,\ }

\usepackage{xcolor}

\hypersetup{colorlinks=true,linkcolor=blue,citecolor=blue,urlcolor=blue} 

\begin{document}

\title{The effect of elastic walls on suspension flow}
\author{Marco Edoardo Rosti, Mehdi Niazi Ardekani, Luca Brandt}
\affiliation{Linn\'{e} Flow Centre and SeRC (Swedish e-Science Research Centre), \\KTH Mechanics, SE 100 44 Stockholm, Sweden}
\date{\today}

\begin{abstract}
We study suspensions of rigid particles in a plane Couette flow with deformable elastic walls. We find that, in the limit of vanishing inertia, the elastic walls induce shear thinning of the suspension flow such that the effective viscosity decreases as the wall deformability increases. This shear-thinning behavior originates from the interactions between rigid particles, soft wall and carrier fluid; an asymmetric wall deformation induces a net lift force acting on the particles which therefore migrate towards the bulk of the channel. Based on our observations, we provide a closure for the suspension viscosity  which can be used to model the rheology of suspensions with arbitrary volume fraction in elastic channels.
\end{abstract}

\maketitle

Understanding how elastic structures interact with fluid flows is a problem attracting a great deal of attention in different fields of science and technology, ranging from biological applications~\cite{grodzinsky_lipshitz_glimcher_1978a, abkarian_lartigue_viallat_2002a, fish_lauder_2006a, greene_banquy_lee_lowrey_yu_israelachvili_2011a, freund_2014a} to energy harvesting~\cite{mckinney_delaurier_1981a, boragno_festa_mazzino_2012a}. In this context we consider a fluid-structure interaction problem particularly relevant to understand biological flows: we study the rheology of suspensions in the presence of walls which are allowed to deform elastically.

The study of rheology is motivated by the many fluids in nature and industrial applications which exhibit a non-Newtonian behavior, \ie a non-linear relation between the shear stress and the shear rate, such as shear thinning, shear thickening, yield stress, thixotropy, shear banding and viscoelastic behaviors. 
The relation between these macroscopic behaviors and the microstructure is often studied assuming suspensions of objects in a Newtonian solvent with dynamic viscosity $\muf$ and density $\rho$. In the simplest case of suspensions of rigid spheres in a Newtonian fluid, Einstein \cite{einstein_1956a} showed in his pioneering work that in the limit of vanishing inertia and for dilute suspensions (\ie $\Phi \rightarrow 0$), the relative increase in effective viscosity $\mue$  is a linear function of the particle volume fraction $\Phi$. However, at present there is no theory that allows us to calculate $\mue$ for any given $\Phi$, thus different empirical formulas have been proposed to provide a good description to the existing experimental and numerical results~\cite{ferrini_ercolani_de-cindio_nicodemo_nicolais_ranaudo_1979a, singh_nott_2003a, kulkarni_morris_2008a}. Among those, we consider here the Eilers formula~\cite{stickel_powell_2005a, mewis_wagner_2012a},
\begin{equation}
\frac{\mu}{\muf} = \left[ 1 + B \frac{\Phi}{1-\Phi/\Phi_m}\right]^2 \/,
\label{eq:Eilers}
\end{equation}
which well fits experimental and numerical data~\cite{zarraga_hill_leighton-jr_2000a,singh_nott_2003a} for both low and high values of $\Phi$, up to about $0.6$. In the expression above, $\Phi_m$ is the geometrical maximum packing fraction and $B$ a constant; fits to the data are usually obtained for $\Phi_m=0.58 - 0.63$ and $B=1.25 - 1.7$.

\begin{figure}
\centering
\includegraphics[width = 0.6\textwidth]{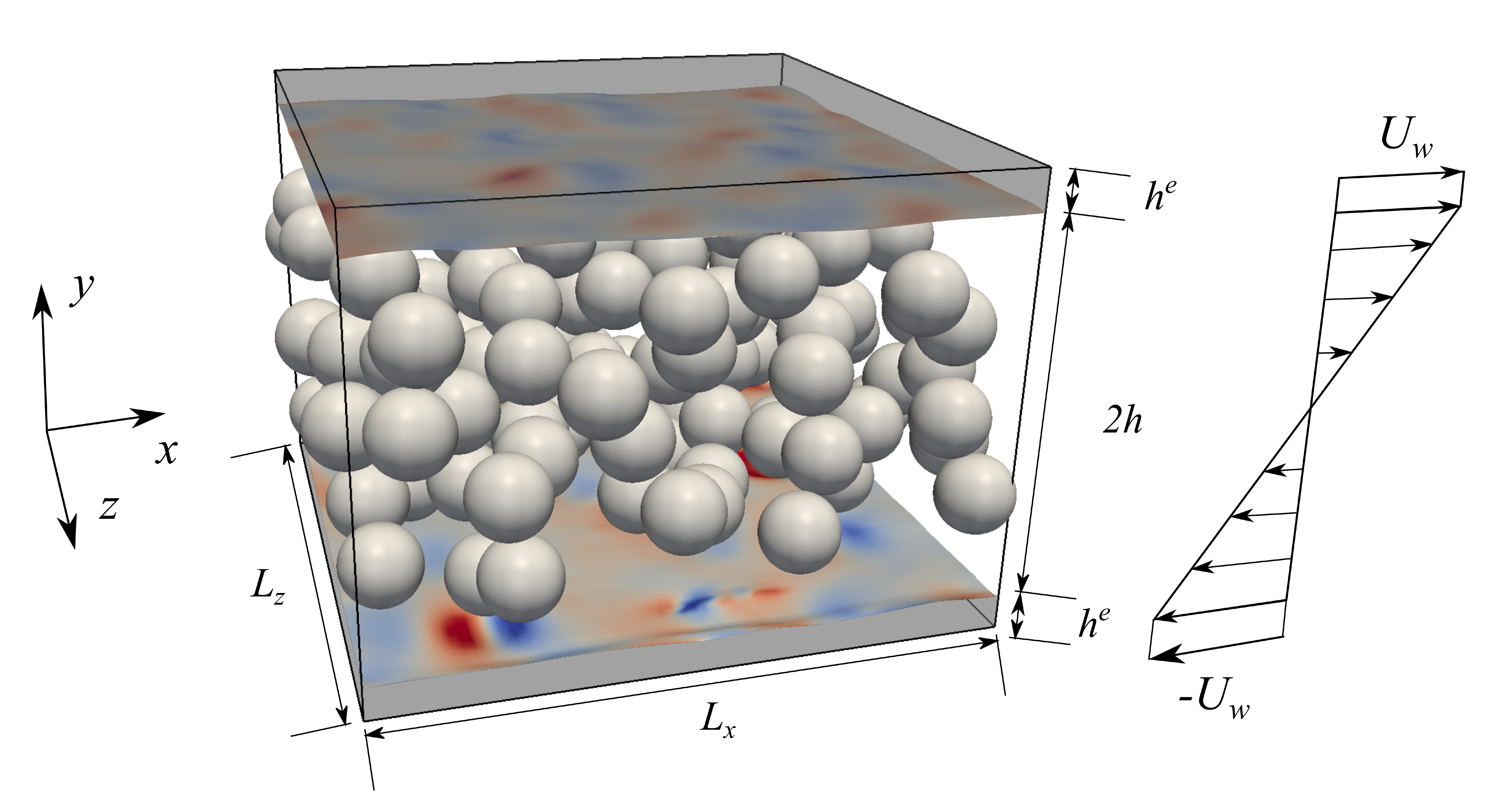} 
\caption{Visualisation of the particles suspension in a Couette flow with elastic walls, with a sketch of the relevant geometrical parameters.}
\label{fig:3d}
\end{figure}

In this letter we add a further complexity to this problem, one that is particularly relevant to understand rheology of biological flows: we allow the walls to deform elastically. The problem of a single object interacting with a soft wall has been the object of several recent works~\cite{mani_gopinath_mahadevan_2012a, beaucourt_biben_misbah_2004a, skotheim_mahadevan_2004a, salez_mahadevan_2015a, saintyves_jules_salez_mahadevan_2016a, rallabandi_saintyves_jules_salez_schoenecker_mahadevan_stone_2017a, davies_debarre_el-amri_verdier_richter_bureau_2018a, rallabandi_oppenheimer_zion_stone_2018a}; on the other hand, here we focus for the first time on particle suspensions interacting with soft walls. In particular, we model the walls as two viscoelastic layers with an elastic shear--modulus $G$ and viscosity $\mum$. Thus, we introduce two new dimensionless parameters in the problem: the capillary number $\Ca\equiv \muf\gdot/G$ and the solid to fluid viscosity ratio $K\equiv \mum/\muf$, the latter fixed to $1$ for simplicity. We aim to describe for the first time the non linear effects of the wall elasticity on the rheology of a suspension. Furthermore, we will provide a simple approach able to accurately model the presence of deformable walls, without the necessity to solve the complex nonlinear interactions between the multiphase flow and the structure dynamics.

\begin{figure}
\centering
\includegraphics[width = 0.49\textwidth]{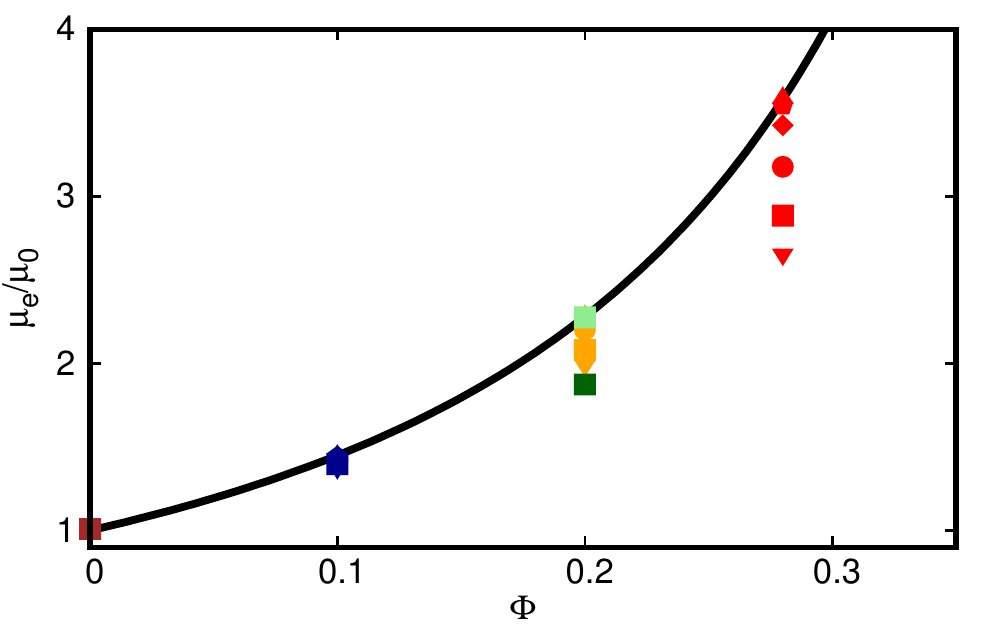}
\includegraphics[width = 0.49\textwidth]{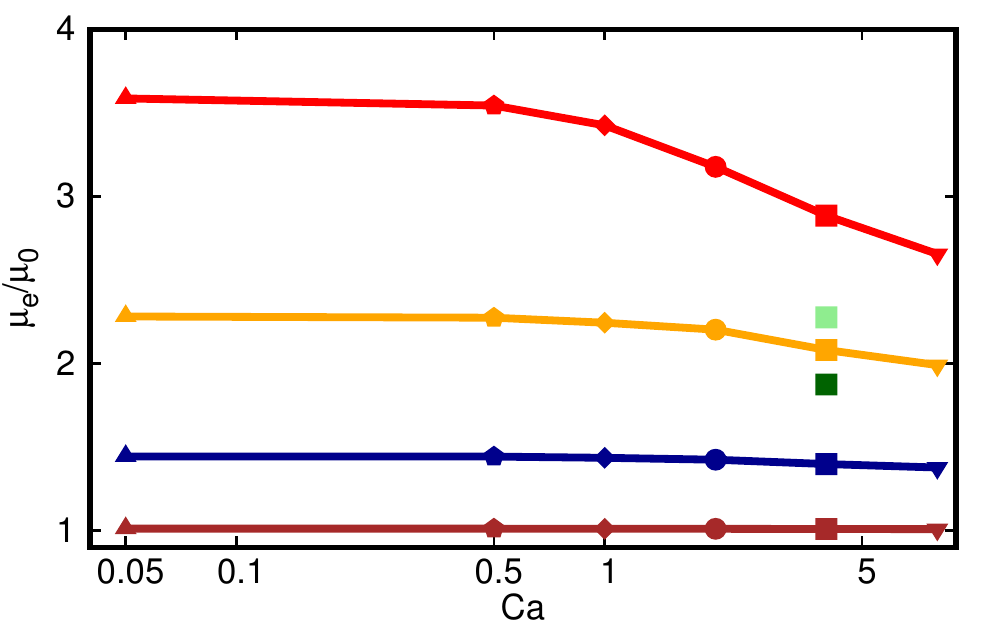} 
\caption{Effective viscosity of the suspension $\mue$ as a function of the volume fraction $\Phi$ (left) and capillary number $\Ca$ (right). The effective viscosity is normalized with the fluid viscosity $\muf$. The brown, blue, orange and red colors are used for volume fractions $\Phi$ equal to $0.0016$, $0.1$, $0.2$ and $0.29$, while the symbols $\triangle$, $\pentagon$, $\lozenge$, $\Circle$, $\square$ and $\triangledown$ for capillary numbers $\Ca$ equal to $0.05$, $0.5$, $1$, $2$, $4$ and $8$. All the previous symbols and colors are used for an elastic layer of thickness $h^e=R$; the light and dark green square represents the results for $\Phi=0.2$, $\Ca=4$ and $h^e=0.5R$ and $1.5R$, respectively. The black line in the left panel is the Eilers formula, \Eq{eq:Eilers}, with $\Phi_m=0.6$ and $B=1.7$.}
\label{fig:visc}
\end{figure}

To address this problem, we perform direct numerical simulations of suspensions of rigid spheres, simulated by an immersed boundary method \cite{izbassarov_rosti_niazi-ardekani_sarabian_hormozi_brandt_tammisola_2018a}, flowing in a plane channel Couette flow. Two viscoelastic layers are attached to the moving rigid walls, simulated with a pseudo volume of fluid approach \cite{rosti_brandt_2017a}, see \Fig{fig:3d} for a sketch of the geometry considered \cite{note1}. We cover a wide range of volume fractions $\Phi$ (up to $30\%$) and capillary numbers $\Ca$ (up to $8$) and calculate the effective viscosity of the suspension $\mue = \mathcal{F} \left( \Phi, \Ca \right)$. This is displayed in \Fig{fig:visc} for different volume fractions $\Phi$ and for various level of wall elasticity, \ie for various capillary numbers $\Ca$. Also, three different thicknesses $h^e$ of the elastic walls are considered: $0.5R$, $R$ and $1.5R$. We observe that, the suspension viscosity increases with the volume fraction $\Phi$ but decreases with the wall deformability (increasing capillary number). The effect of the wall deformability  is more pronounced as the volume fraction increases, where we observe larger differences from the reference case (rigid walls), indicated by the Eilers fit in the figure (solid black line). In particular, we find a reduction of up to $25\%$ in the effective viscosity at $\Phi\approx30\%$ for the largest capillary number considered in this study. Note that, the reduction of the suspension viscosity with the capillary number can be interpreted as a shear-thinning rheological behavior of the system, see \Fig{fig:visc}, where the effective viscosity $\mue$ is shown as a function of $\Ca$. Interestingly, a similar behaviour was observed in Ref.~\cite{rosti_brandt_mitra_2018a} for suspensions of deformable particles. It is worth noticing that, the presence of the elastic wall has no effect on the fluid rheology in the absence of particles, \ie  $\mu_e=\mu_0$ when $\Phi=0$. Thus, the observed shear thinning  is the result of the interaction of the elastic walls and the particles. This  behaviour is affected by the thickness of the elastic layer $h^e$: when the size of the layer is increased, the effect on the suspension is enhanced and the effective viscosity further reduces.

\begin{figure}
\centering
\includegraphics[width = 0.49\textwidth]{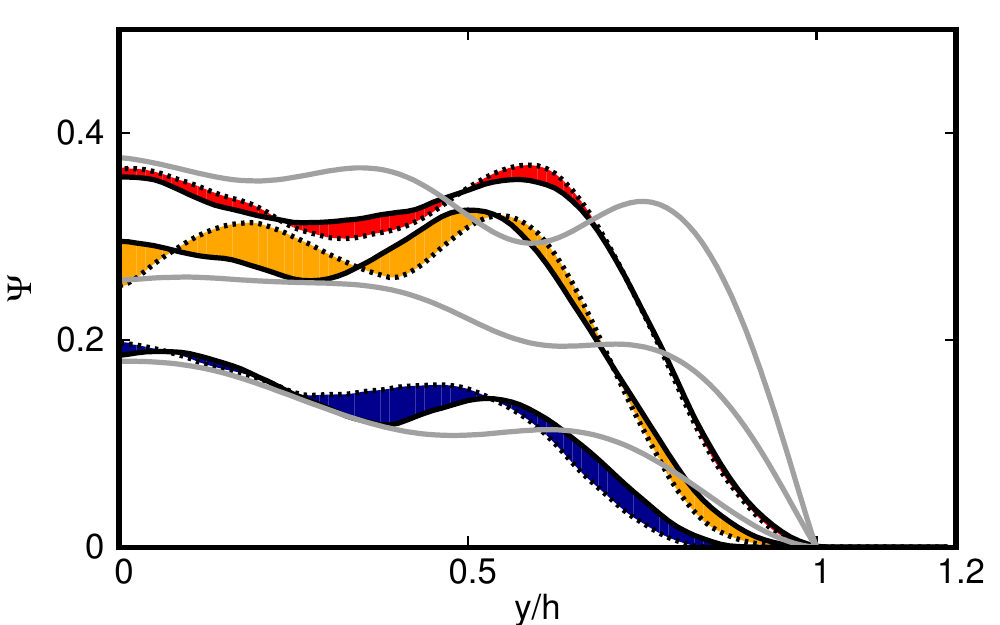}
\includegraphics[width = 0.49\textwidth]{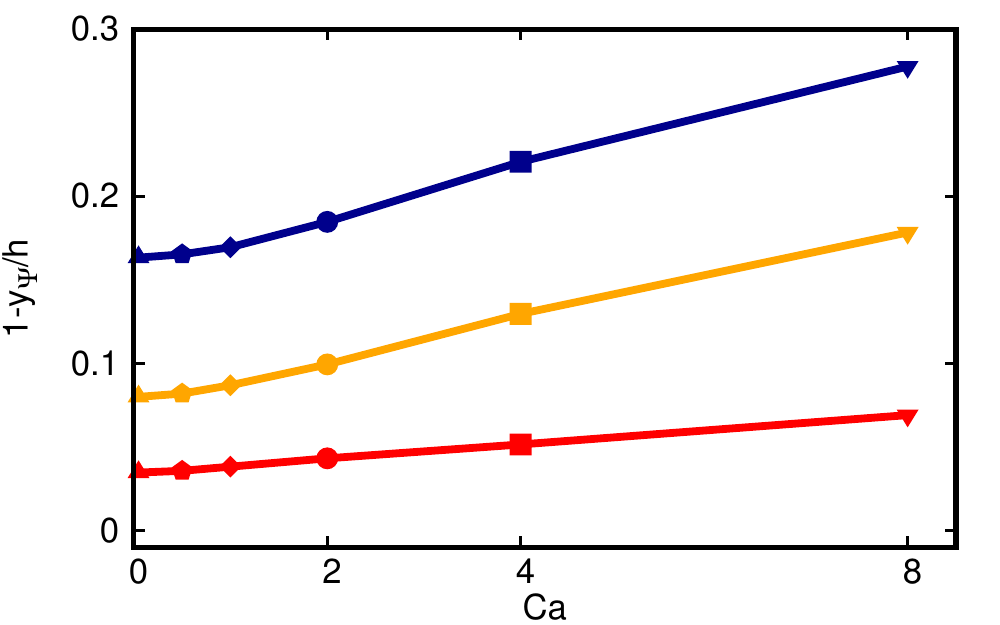} 
\caption{(left) Mean particles concentration $\Psi$ as a function of $y$. The shaded areas represent the variation due to the capillary number $\Ca$, with the black solid and dashed lines representing the lowest and highest values of $\Ca$; the grey solid lines are the data for rigid walls. (right) Distance from the wall $h-y$ at which the mean particles concentration is equal to $0.01$ of the nominal value as a function of the capillary number $\Ca$.}
\label{fig:conc}
\end{figure}
To understand the mechanism that generates the shear-thinning behavior of the suspension, we study the mean particle concentration $\Psi(y)$ across the channel, shown in the left panel of \Fig{fig:conc}. The particles concentration $\Psi$ is null inside the elastic layer ($y \geq h$), rapidly grows in the near wall region ($h/2 \lesssim y \lesssim h$) and finally reaches an approximately uniform value in the middle of the channel ($h \lesssim h/2$). Unlike the case of rigid walls (grey line), on average there are no particles in contact to the deformable wall as these are lifted towards the channel center. As the total volume fraction $\Phi$ increases, the concentration distribution grows faster close to the wall and reaches higher values at the bulk of the channel to ensure the imposed total volume fraction $\Phi$. The effect of the wall elasticity is shown by the shaded colored areas in the graph: as $\Ca$ increases, the particles are displaced further away from the wall and concentrate more in the bulk of the channel. This effect is present for all the volume fractions $\Phi$ and capillary numbers $\Ca$ that we studied, but is more evident at low volume fractions than at high ones due to the larger available space for the particles to migrate. Note that, despite this, the effect on the effective viscosity of the suspension increases with the volume fraction, as shown in \Fig{fig:visc}. The mean displacement of the particles from the wall is quantified in the right panel of \Fig{fig:conc}, where the wall-normal distance ($h-y$) at which the concentration $\Psi$ is equal to $1\%$ of the nominal value is reported as a function of the capillary number $\Ca$.
The wall-normal distance where $\Psi=1\%$ increases for all the volume fractions with the capillary number $\Ca$, \ie as the wall becomes more deformable, and decreases for increasing total volume fraction $\Phi$.

\begin{figure}[b]
\centering
\includegraphics[width = 0.49\textwidth]{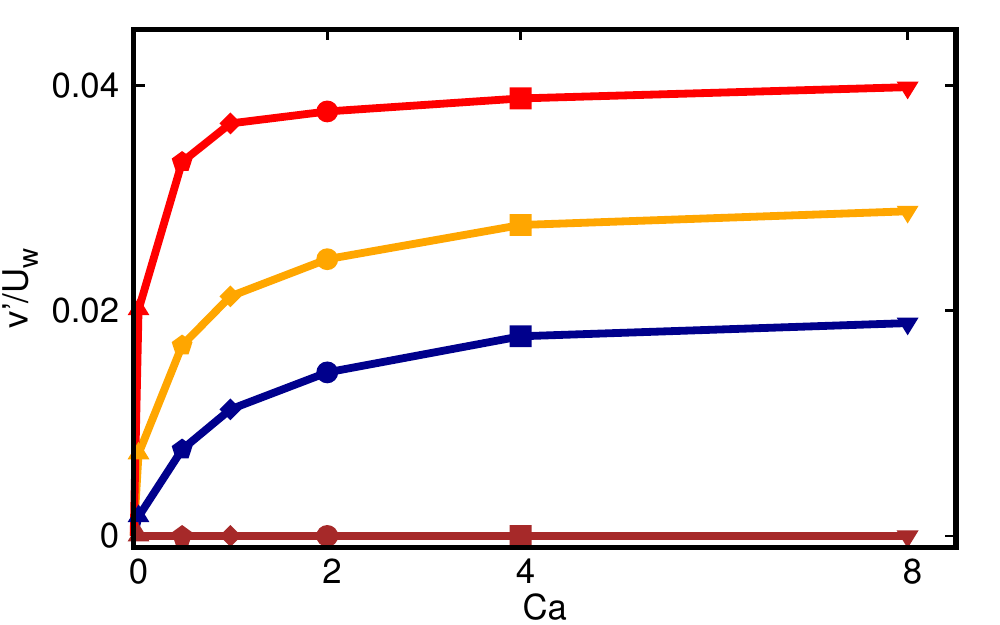} 
\caption{Root mean square of the wall-normal velocity fluctuations $v'$ at the interface between the fluid and deformable wall $y=h$ as a function of the capillary number $\Ca$.}
\label{fig:vel}
\end{figure}
\begin{figure}
\centering
\includegraphics[width = 0.6\textwidth]{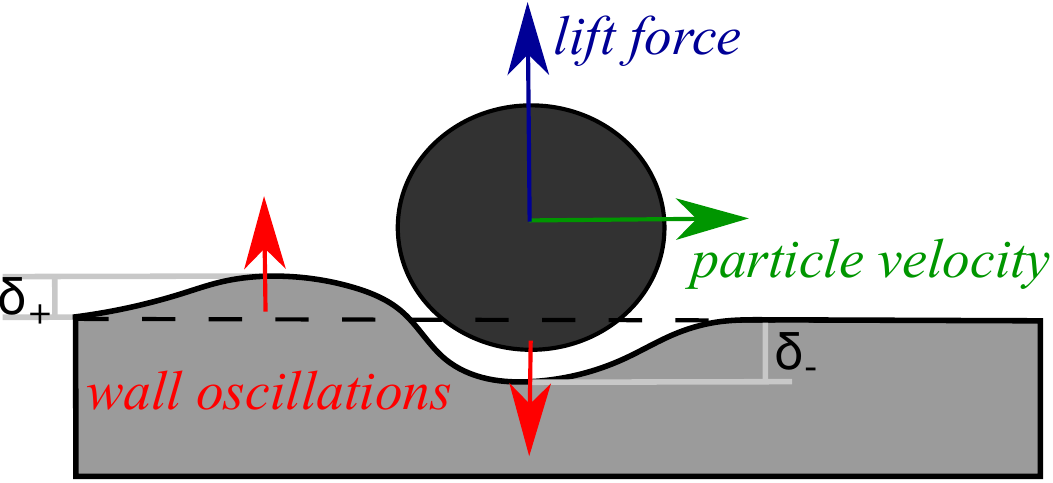} 
\caption{Sketch of the mechanism of the lift generation induced by the interaction of a moving sphere and a rigid particle.}
\label{fig:sketch}
\end{figure}

The displacement of the particles from the walls to the center of the channel can be related to the non-zero wall-normal velocity fluctuations $v'$ at the elastic walls ($y=h$), although the mean wall-normal velocity is zero for incompressible materials. These fluctuations are shown in \Fig{fig:vel}, where $v'$ is reported as a function of the capillary number $\Ca$ for all the volume fractions $\Phi$ considered. We observe that $v'$ grows from zero as the capillary number is increased, \ie the wall is allowed to deform; the fluctuation growth rate is high for low capillary numbers before we observe a tendency to saturate. Also, the decrease of the growth rate at high capillary numbers is faster for high volume factions, in accordance with what already observed in \Fig{fig:conc}. The wall-normal velocity fluctuations are induced by the deformation of the walls due to the interaction with the rigid particles, as sketched in \Fig{fig:sketch}. 
A particle approaching the elastic wall deforms it generating a wall-normal flow; in addition, the particle moves along the wall due to the applied shear. The combination of these two effects induces an asymmetric wall deformation which leads to the generation of a net lift force acting on the particle, which therefore migrates away from the wall. These results are consistent with and confirm previous theoretical works \cite{salez_mahadevan_2015a, rallabandi_saintyves_jules_salez_schoenecker_mahadevan_stone_2017a} which considered the deformations induced by a single rigid sphere translating parallel to a soft wall in a viscous fluid;  in these previous studies, lubrication theory is shown to predict a lift force acting on the object, as observed experimentally \cite{saintyves_jules_salez_mahadevan_2016a}.

\begin{figure}[b]
\centering
\includegraphics[width = 0.49\textwidth]{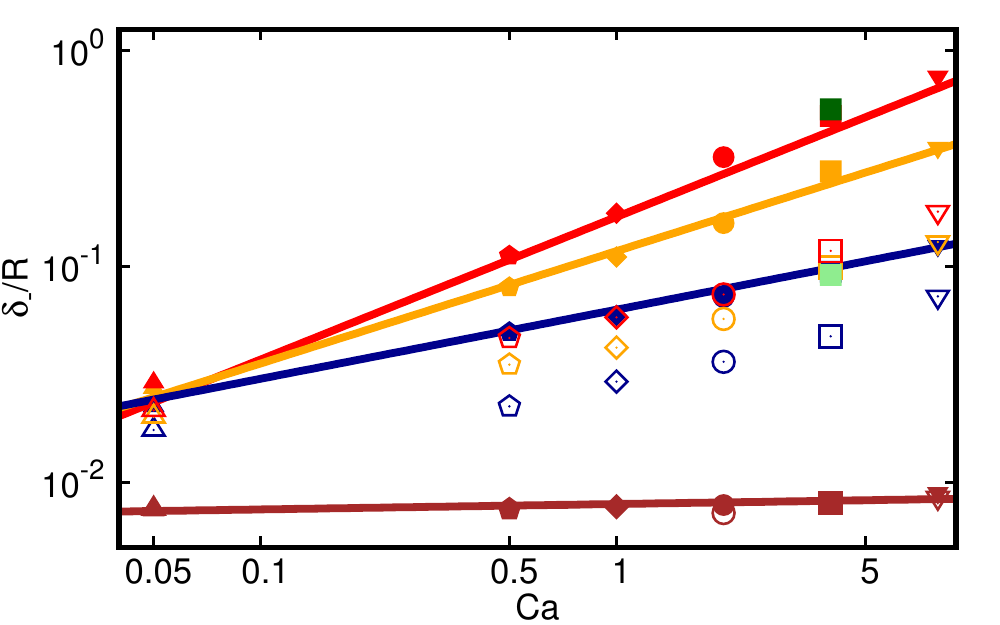} 
\includegraphics[width = 0.49\textwidth]{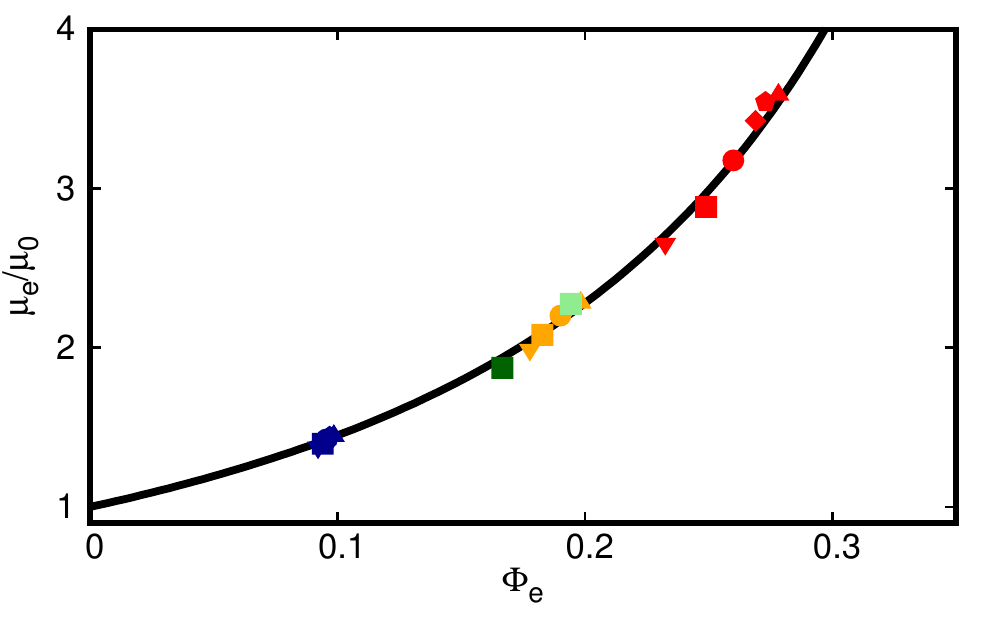} 
\caption{(left) Mean wall deformations $\delta_\pm$ as a function of the capillary number $\Ca$. The solid and open points are the negative $\delta_-$ and positive $\delta_+$ wall deformations, respectively, while the solid lines are fits to our data in the form of \Eq{eq:fit1}. (right) Normalized effective viscosity $\mue$ as a function of the effective volume fraction $\phie$ computed from \Eq{eq:fit2}. The black solid line is the Eilers formula reported in \Eq{eq:Eilers}.}
\label{fig:fit}
\end{figure}

Finally, we propose a model to easily include the effect of the elastic walls in the rheological description of suspensions. As mentioned above, $\mu_e$ is in general a non-linear function of the total volume fraction $\Phi$ and the capillary number of the walls $\Ca$, \ie $\mu_e=\mathcal{F} \left( \Phi, \Ca \right)$. Below, we show that $\mu_e$ can be written as a function of a single variable $\Phi_e$, \ie $\mu_e=\mathcal{G} \left(\Phi_e \right)$. Also, we will demonstrate that the nonlinear function $\mathcal{G}$ can be properly described by the Eilers fit reported in \Eq{eq:Eilers}, similarly to the standard case of flow over rigid walls. The effective volume fraction $\Phi_e$ takes into account the fact that the particles feel a reduced confinement effect due to the wall flexibility: indeed, they can move beyond $y=h$ by inducing wall deformations. Thus, we introduce a reduced effective volume fraction, obtained by increasing the total volume available to the particles by a layer of thickness $\delta_-$, which is the amount of the penetration of the particles into the walls, or alternatively, the amplitude of the average negative wall deformation. We plot in the left panel of \Fig{fig:fit} the normalized amplitude of wall deformation $\delta_-$, which grows with both the wall elasticity $\Ca$ and the volume fraction $\Phi$. A simple fit of our data (shown with the solid lines in the plot) provides the following expression for $\delta_-$ as function of the volume fraction $\Phi$, the capillary number $\Ca$ and the elastic layer thickness $h^e$
\begin{equation}
\label{eq:fit1}
\frac{\delta_-}{R} \approx \Phi^{c_1} ~ \Ca^{c_2} \left( \frac{h^e}{R} \right)^{c_3},
\end{equation}
being $c_i$ fitting coefficients,  $c_1 \approx 0.9$, $c_3 \approx 1.5$ and $c_2 \approx 0.32$ for $\Phi=0.1$, $0.52$ for $\Phi=0.2$ and $0.66$ for $\Phi=0.29$. Note that, \Fig{fig:fit} reports also $\delta_+$, the amplitude of the positive wall deformation, which is always smaller than $\delta_-$ due to the aforementioned symmetry breaking at the deformable wall interface.
Once $\delta_-$ is known, one can compute the effective volume fraction as
\begin{equation}
\label{eq:fit2}
\Phi_e = \mathcal{N} \frac{\frac{4}{3} \pi R^3}{\mathcal{V}^e_{tot}},
\end{equation}
where $\mathcal{V}^e_{tot}$ is the sum of the total volume originally available to the particles $\mathcal{V}_{tot}=2h~L_x~L_z$ and the increase due to the wall elasticity $\mathcal{V}^e=\delta~L_x~L_z$, \ie $\mathcal{V}^e_{tot}=\mathcal{V}_{tot}+\mathcal{V}^e = 2\left( h + \delta_- \right)~L_x~L_z$. The right panel in \Fig{fig:fit} shows again the effective viscosity $\mue$ now as a function of the effective volume fraction $\phie$. We observe that all the cases at different volume fractions $\Phi$, capillary numbers $\Ca$ and wall thickness $h^e$ collapse onto a single curve, which is well described by the Eilers formula. Thus, by using \Eq{eq:fit1} and \Eq{eq:Eilers} we are able to effectively predict the suspension effective viscosity $\mue$ in the presence of elastic walls, once the volume fraction $\Phi$, the capillary number $\Ca$ and the thickness of the elastic layers $h^e$ are known. Also, we explain the largest variations of the effective viscosity $\mue$ at higher volume fractions by the large sensitivity of $\mue$ to variations of $\Phi$ at higher concentrations (slope of the Eilers curve).

In conclusion, we have explored the rheological behaviour of a suspension of rigid particles in an elastic channel, for a wide range of solid volume fractions and wall elasticities. The main results of our analysis are the identification of a shear-thinning behavior of the suspension induced by the interaction of the elastic walls and the rigid particles, and its explanation in terms of a lift force acting on the particles which induces their migration towards the bulk of the channel. The lift force mechanism is robust enough to be effective also in relatively dense suspensions, whose behaviour is usually mainly determined by inter-particle interactions. Based on our observations and a simple mechanical model, we also provide a closure to effectively predict the rheological properties of suspensions in the presence of elastic walls. Our results extend to deformable walls the idea of using simple empirical fits, originally valid for inertialess suspensions of rigid spheres, such as the Eilers formula, to predict the rheology of suspensions with additional complexity embedded in the definition of an effective volume fraction, as previously done in Refs. \cite{picano_breugem_mitra_brandt_2013a}, \cite{mueller_llewellin_mader_2010a} and \cite{rosti_brandt_mitra_2018a} for the cases of inertial effects, particles shape and deformability, respectively. The applicability of this scaling confirms that viscous dissipation is still the dominant mechanism at work in these flows.

\section*{Acknowledgment}
\noindent The authors were supported by the ERC-2013-CoG-616186 TRITOS and by the VR 2014-5001 and acknowledge the computer time provided by SNIC (Swedish National Infrastructure for Computing).


\end{document}